\documentclass[twocolumn,showpacs,amsmath,amssymb]{revtex4}

\usepackage{graphicx}
\usepackage{dcolumn}
\usepackage{bm}

\begin{document}
\title{Physics of collisionless reconnection in a
stressed X-point collapse}
\author{D. Tsiklauri}
\author{T. Haruki}

\affiliation{Joule Physics Laboratory,
Institute for Materials Research, University of Salford, Manchester, M5 4WT,
United Kingdom}
\date{\today}

\begin{abstract}
Recently, magnetic reconnection during collisionless, stressed, X-point collapse was studied 
using kinetic, 2.5D, fully electromagnetic, relativistic 
Particle-in-Cell numerical code [D. Tsiklauri and T. 
Haruki, Phys. Plasmas {\bf 14}, 112905 (2007)]. Here we finalise the investigation of this topic by addressing
key outstanding physical questions: (i) which term in the generalised Ohm's law is responsible
for the generation of the reconnection electric field?
(ii) how does the time evolution of the reconnected flux vary with the ion-electron mass ratio?
(iii) what is the exact energy budget of the reconnection process, i.e. in which proportion initial 
(mostly magnetic) energy is converted into other forms of energy?
(iv) are there any anisotropies in the velocity distribution of the accelerated particles?
It has been established here that: (i) reconnection electric field is generated by 
the electron pressure tensor off-diagonal terms, resembling to the case of tearing unstable
Harris current sheet studied by the GEM reconnection challenge;
(ii) For $m_i / m_e \gg 1$ the time evolution of the reconnected 
flux is independent of ion-electron mass ratio;
also, in  the case of $m_i / m_e = 1$ we show that reconnection proceeds slowly as the
Hall term is zero; when $m_i / m_e \gg 1$ (i.e. the Hall term is non-zero) reconnection is fast and
we conjecture that this is due to magnetic field being frozen into electron fluid, which
moves significantly faster than ion fluid;
(iii) within one Alfv\'en time, somewhat less than half ($\sim 40$\%) of the 
initial total (roughly magnetic) energy is converted into the kinetic 
energy of electrons, and somewhat more than  half ($\sim 60$\%) into kinetic 
energy of ions (similar to solar flare observations);
(iv) in the {\it strongly} stressed X-point case, in about one Alfv\'en time, a full 
isotropy in all three spatial directions of the velocity distribution is seen for super-thermal 
electrons (also commensurate to solar flare observations).
\end{abstract}	


\maketitle

\section{Introduction}

In many astrophysical or laboratory plasma situations (a) plasma beta is small, indicative of large
amounts of energy stored in a form of magnetic field and (b) there is a need to 
explain or provide plasma heating, as well
as  plasma particle acceleration. It is believed that in such situations magnetic reconnection, i.e. change
of connectivity of magnetic field lines that penetrate the 
plasma, can serve as one of the important possible mechanisms.
There are different types of magnetic reconnection. One of the key descriptors is plasma collisionality,
i.e. if plasma is collisional then magnetic resistivity, $\eta$ (or more specifically $\eta {\vec j}$ term 
in the generalised Ohm's law) is responsible for breaking the frozen-in condition
(enabling field line connectivity change).
However, if plasma is collisionless, then other terms in the generalised Ohm's law may be more important.
In this context, it is instructive to look at typical spatial 
scales. Let us consider an example of solar coronal plasma.
Typical width of a Sweet-Parker current sheet is given by $\delta=S^{-1/2} L$ \cite{biskamp} (p. 54).
Fixing coronal temperature at 1 MK, Coulomb logarithm at 18.0948, the Lundquist number (using Spitzer
resistivity from \citet{NRLF}, p. 30) is $5.37933 \times 10^{12}$, which for $L=10$ Mm yields, $\delta=4.31157$ m.
Another way of looking at $\delta$ is associating it also with the
resistive length scale via Alfv\'en time scale $\tau_A = L / V_A$, where $V_A$ is the Alfv\'en speed 
($\approx 1$ Mm s$^{-1}$): i.e. $\delta=S^{-1/2} L=\eta \tau_A / \mu_0$.
Typical scale associated with the Hall term in the generalised Ohm's law at which deviation from
 electron-ion coupled dynamics is observed is, $c/\omega_{pi} =  7.20064$ m (ion inertial length). Here 
particle density of $n=10^{15}$ m$^{-3}$ is used.
The fact that $\delta / (c/\omega_{pi}) \leq 1 $  points to a necessity of
 going beyond single fluid MHD approximation.
The Hall term itself cannot break the frozen-in condition, its inclusion into consideration 
ensures that the magnetic field is frozen into electron fluid. Below, we shall use this argument to conjecture
why reconnection is fast when the Hall term is included. 
Importance of different terms in the generalised Ohm's law is usually inferred 
by comparing the spatial scales associated with them to the resistive length scale $\delta$.
E.g. one of the other noteworthy scales is
$c/\omega_{pe} =   0.16804$ m (electron inertial length) on which the electron inertia term operates.
When $\delta / (c/\omega_{pe}) < 1$ then electron inertia would dominate over resistive diffusion \cite{biskamp} (p. 200).
As can be seen from the above estimates  $\delta / (c/\omega_{pe}) = 25.6576 \gg 1 $ in the solar
corona, thus electron inertia effects seem to be negligible.
However, electron inertia in contrary to the Hall term (as well as the electron pressure tensor) 
can break the frozen-in condition and thus change the magnetic field connectivity.
Because, of the fact that with increase of $T$ (hot plasmas), 
$\delta$ gets progressively smaller thus effects other than the resistivity should be included.
Indeed, collisionless (non-resistive) reconnection has recently attracted considerable
attention (see \citet{bp07} for a review). 

One of the first studies of magnetic reconnection is stressed, X-point collapse  \cite{d53} (also see Chap. 7.1 in
\citet{pf00}). The latter was using resistive MHD approach. We recently 
revisited the problem in the regime of collisionless reconnection \cite{th07}. 
In Ref. \cite{th07} we studied the magnetic reconnection during collisionless, stressed, X-point collapse  
using kinetic, 2.5D, fully electromagnetic, relativistic Particle-in-Cell numerical code. 
We considered  two cases of weakly and strongly stressed X-point collapse.
Where descriptors weakly and strongly 
refer to 20 \% and 124 \% unidirectional spatial compression of the X-point, respectively. 
Amongst other interesting outcomes, we established that  
within about one Alfv\'en time,  2\%  and 20\% of the initial 
magnetic energy can be  converted into heat and accelerated particle energy in the 
cases of weak and strong stress, respectively. However, open questions remained:
(i) which term in the generalised Ohm's law is responsible
for the generation of the reconnection electric field?
(ii) how does the time evolution of the reconnected flux vary with the ion-electron mass ratio?
(iii) what is the exact energy budget of the reconnection process, i.e. in which proportion initial 
(mostly magnetic) energy is converted into other forms of energy?
(iv) are there any anisotropies in the velocity distribution of the accelerated particles?
Here we finalise the study of magnetic, collisionless reconnection of a stressed X-point by
providing answers to these questions.

\section{The Model}

The numerical code used here is 2.5D, 
relativistic, fully electromagnetic PIC code, with the initial conditions 
the same as in our previous work \cite{th07}. For completeness we re-iterate
key points: 
Magnetic field configuration is an X-point without a guide-field
\begin{equation}
  (B_x, B_y, B_z) = \frac{B_0}{L} (y, \alpha^2 x, 0),
\label{eq:init_b}
\end{equation}
where $B_0$ is magnetic field intensity at the distance $L$ from the X-point ($L$ is the global system scale).
$\alpha$ is the 
{\it stress parameter}, which prescribes 
the initial strength of magnetic pressure that collapses the system, due to lack of restoring force \cite{pf00}.
Using $\mu_0 \vec{j} = \nabla \times \vec{B}$, 
a uniform current is imposed in the $z$ direction, 
\begin{equation}
  j_z = \frac{B_0}{\mu_0 L} (\alpha^2 - 1).
\label{eq:init_j}
\end{equation}
Electrons and ions  have uniform spatial, and Maxwellian velocity distributions 
throughout the system.
For $\alpha = 1$, magnetic field 
geometry is completely symmetric, $j_z$ current zero
 (see Eqs.~(\ref{eq:init_b})-(\ref{eq:init_j})), and thus, such
magnetic configuration is stable.
For $\alpha > 1$,  stressed X-point starts collapse in the $x$ direction 
because of the absence of a restoring force, causing time-transient 
magnetic reconnection.
The main parameters of the standard simulation 
model are as follows.
The length of the system in 
two dimensions is $L_x = L_y = 400 \Delta$ (this is excluding so-called ghost cells), 
where $\Delta = 1$ is the simulation grid size corresponding 
to electron Debye length, $\lambda_D = v_{te} / \omega_{pe} = 1 \Delta$ ($v_{te}$ is 
electron thermal velocity and $\omega_{pe}$ is electron plasma frequency).
The global reconnection scale is set $L = 200 \Delta$.
The number density is fixed at $n_0 = 100$ electron-ion pairs per cell.
Hence the total number is $1.6 \times 10^7$ pairs.
The simulation time step is $\omega_{pe} \Delta t = 0.05$.
Ion-to-electron mass ratio is fixed at 
$m_i / m_e = 100$ (which is varied in Fig. 2 (subsection III.B) only).
The electron thermal velocity to speed of light ratio is $v_{te} / c = 0.1$.
The electron and ion skin depths are $c / \omega_{pe} = 10 \Delta$ and $c / \omega_{pi} = 100 \Delta$, respectively. 
The electron cyclotron frequency to plasma 
frequency ratio is $\omega_{ce} / \omega_{pe} = 1.0$ for magnetic field intensity, $B = B_0$.
This ratio is 
close to unity in the solar 
corona, while it is much 
bigger than unity in the Earth magnetosphere.
The electron and ion Larmor 
radii are $v_{te} / \omega_{ce} = 1 \Delta$ and $v_{ti} / \omega_{ci} = 10 \Delta$, where $v_{ti}$ 
is the ion thermal velocity. 
Initial temperatures of electrons and ions are initially 
set the same, $T_e = T_i$.
At the boundary ($B = B_0$ at the distance $L$ from the X-point), the plasma $\beta = 0.02$ and 
Alfv\'en velocity, $V_{A0} / c = 0.1$.
Naturally these vary across the 
simulation box as the background magnetic field is a function of $x$ and $y$.

The boundary conditions on EM-fields are zero-gradient and 
also, tangential component of electric field 
 was forced to zero, while normal component of magnetic field was kept
constant, both at the boundary.
  This ensures that there is no change in magnetic flux through the simulation box,
  i.e. the system is isolated. 
When colliding with boundaries particles are reflected. Thus our
boundary conditions ensure there is no magnetic influx or 
 mass transport across the boundaries. It has been also confirmed that
the total energy in the system is conserved during the simulations to a good accuracy.

\section{Results}

Before we address the outstanding questions, we refer reader to
\citet{th07} for a more detailed description of dynamics of EM-fields, currents, and particles. 
It is our intention to focus on the outstanding questions here.
In brief, the previous results can be summarised as follows:
when $\alpha > 1$,  the stressed X-point collapses in the $x$ direction 
due to the absence of a restoring force, and hence time-transient 
magnetic reconnection occurs. The fast reconnection regime is achieved.
Initially uniform out-of-plane current becomes localised,
peaking at many times its initial value in a time transient manner.
Also, out-of-plane quadruple
magnetic field is generated.
High energy part of the electron distribution
function exhibits a power-law behaviour. Sizable fraction of
initial magnetic energy is converted into other forms of energy.

\subsection{Source of the reconnection electric field}

\begin{figure}
\includegraphics[scale = 0.5]{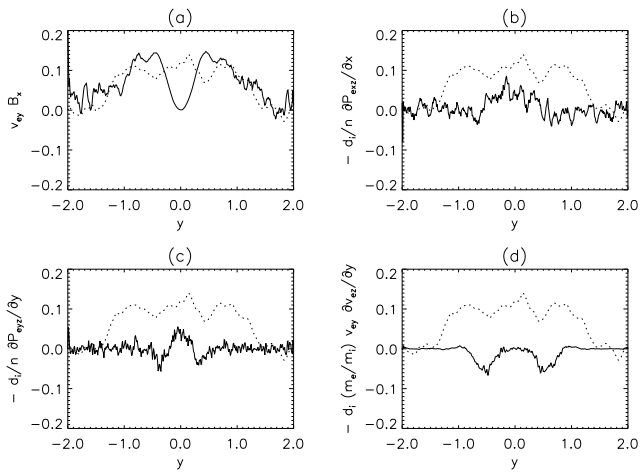}
\caption {\label{fig:ohms_law} Line plots of different terms in the
generalised Ohm's law along 
 $y$ direction, in $x = 0$, at $\omega_{pe} t = 170$ (time-transient reconnection peak) 
 for $\alpha = 1.20$.
 Solid lines in panels (a)-(d) show the different terms as follows:
(a) $v_{ey} B_x$, 
(b) $- (d_i / n) \partial P_{exz} / \partial x$, 
(c) $- (d_i / n) \partial P_{eyz} / \partial y$ and (d) $- d_i (m_e / m_i) v_{ey} \partial v_{ez} / \partial y$. 
The reconnection electric field $E_z(0,y)$ (normalised to $V_{A0} B_0$) is 
shown with the dashed line in all four panels. Here $y$ is normalised to $c/\omega_{pi}$, and thus varies between 
$-2<y<2$.}
\end{figure}

In order to understand details of the collisionless reconnection process, we now
focus on the question: which term in the generalized Ohm's law is responsible
for the generation of the reconnection electric field? We adopt an appoach used by
\citet{pritchett01}. The generalized Ohm's law
 can be written as (e.g.  \cite{bp07} p. 108)
\begin{equation}
  \vec{E} = - \vec{v}_e \times \vec{B} 
            - \frac{\nabla \cdot \vec{P}_e}{n_e e}
            - \frac{m_e}{e} (\frac{\partial \vec{v}_e}{\partial t} 
            + \left( \vec{v}_e \cdot \nabla) \vec{v}_e \right), 
\label{eq:ohms_law}
\end{equation}
where $\vec{E}$ and $\vec{B}$ are electric and magnetic fields, $\vec{v}$ is plasma 
 velocity, $\vec{P}$ is pressure tensor ($3 \times 3$ matrix), $n$ is 
plasma number density, $m$ is  mass and $e$ is electric charge.
The subscript $e$  refers to an electron.
Normalising space coordinates by global reconnection scale 
$L$, fluid velocity by Alfv\'en speed $V_A$, time by 
Alfv\'en transit time $\tau_A (= L / V_A)$, 
magnetic field by $B_0$, number density by $n_0$ and 
pressure tensor by $B_0^2 / \mu_0$, a dimensionless 
version of Eq.(\ref{eq:ohms_law}) can be obtained
\begin{eqnarray}
  \vec{E} = - \vec{v}_e \times \vec{B} 
            - d_i \frac{\nabla \cdot \vec{P}_e}{n_e}
            - d_i \frac{m_e}{m_i} (\frac{\partial \vec{v}_e}{\partial t} 
            + \left( \vec{v}_e \cdot \nabla) \vec{v}_e \right), 
\label{eq:nd_ohms_law}
\end{eqnarray}
where $d_i$ is the normalised ion skin depth ($d_i=c/\omega_{pi}L$).
Note that strictly speaking we should have used tildes in Eq.(\ref{eq:nd_ohms_law})
to denote dimensionless quantities, but we omit them for brevity.

Let us focus on the out-of-plane component of the electric field $E_z$ at the magnetic null, 
which is a measure of the reconnection rate. It is given by,
\begin{eqnarray}
  E_z &=& - (v_{ex} B_y - v_{ey} B_x)
          - d_i \frac{1}{n} \left( \frac{\partial P_{exz}}{\partial x}
                 + \frac{\partial P_{eyz}}{\partial y} \right) \nonumber \\
      & & - d_i \frac{m_e}{m_i} \left( \frac{\partial v_{ez}}{\partial t}
          + v_{ex} \frac{\partial v_{ez}}{\partial x}
          + v_{ey} \frac{\partial v_{ez}}{\partial y} \right),
\label{eq:nd_ez}
\end{eqnarray}
where $\partial / \partial z = 0$ is assumed because of 2D reconnection model.

The pressure tensor is defined as $P_{ij} = m \int v^{'}_i v^{'}_j f (\vec{r}, \vec{v}, t) d\vec{v}$, 
where $m$ is mass, $v^{'}$ is random velocity, 
the subscript $i$ and $j$ denote the components $x$, $y$ or $z$, 
$f$ is the particle velocity distribution function,  
$\vec{r}$ is position, and $\vec{v}$ is velocity.
In order to get the pressure tensor, number density is calculated first,
from $n(\vec{r}, t) = \int f (\vec{r}, \vec{v}, t) d\vec{v}$.
Mean velocity is also obtained 
via $\vec{V} (\vec{r}, t) = (1 / n) \int \vec{v} f (\vec{r}, \vec{v}, t) d\vec{v}$.
For pressure tensor calculation, the number density, $n$ and 
the mean velocity, $\vec{V} (\vec{r}, t)$, is calculated by 
counting number of individual particles 
per cell
and by computing the average 
velocity in each cell, respectively.
We then estimate the random velocity, $\vec{v^{'}} = \vec{v} - \vec{V}$, which is
used in the above definition of the
pressure tensor $P_{ij}$. In PIC simulations, 
in practise, the summation of $m v^{'}_i v^{'}_j$ over all individual particles is used.

Figure \ref{fig:ohms_law} shows $y$-profiles of different terms in the
generalised Ohm's law at $x = 0$, for $\omega_{pe} t = 170$ (time-transient reconnection peak). 
Here $\alpha = 1.20$.  Solid lines in panels (a)-(d) indicate 
different terms as follows: (a) $v_{ey} B_x$, 
(b) $- (d_i / n) \partial P_{exz} / \partial x$, 
(c) $- (d_i / n) \partial P_{eyz} / \partial y$ and (d) $- d_i (m_e / m_i) v_{ey} \partial v_{ez} / \partial y$. 
The reconnection electric field $E_z(0,y)$  is shown with the dashed line in all four panels.
A boxcar average scheme with a width of 7 mesh points is applied for smoothing data.
The other terms in Eq.~(\ref{eq:nd_ez}) are negligibly small.
Fig.~\ref{fig:ohms_law} is analogous to figure 5 from \citet{pritchett01}.
Fig.~\ref{fig:ohms_law}(a) shows that in all regions {\it except} the magnetic
null, (0,0), contribution to $E_z(0,y)$ from the $\vec v_e \times \vec B$ term
is significant. However, $v_{ey} B_x$ is zero  at the X-point (the magnetic null).
As seen in Fig.~\ref{fig:ohms_law}(b-c), the off-diagonal components of the 
electron pressure tensor are major contributors to $E_z(0,0)$.
The electron inertia term also generates the electric field {\it away} from the X-point
(see Fig.~\ref{fig:ohms_law} (d)). Thus, we conclude that the reconnection electric field is generated by 
the electron pressure tensor off-diagonal terms; and hence the latter are responsible for breaking
the frozen-in condition. A similar conclusion was reached by \citet{pritchett01}. This
coincidence seems somewhat unexpected, because X-point collapse considered here and
onset of tearing instability considered by \citet{pritchett01} are physically different.
Similarity of the cause of breaking of the frozen-in condition in both cases can only point to
a universal nature of this mechanism.

\subsection{Effect of variation of the ion-electron mass ratio
and conjecture of fast reconnection}

\begin{figure}
\includegraphics[scale = 0.5]{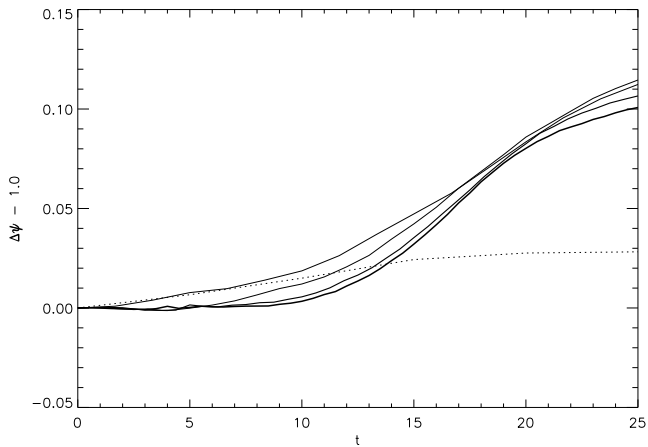}
\caption{\label{fig:mass_depend}
Time evolution of the magnetic flux difference between the O and X points (i.e. amount of reconnected flux).
The solid lines with progressively increasing thickness
 show cases of $m_i / m_e = 9, 25, 64$ and $100$, respectively.
The dotted line shows electron-positron plasma case ($m_i / m_e = 1$).
The magnetic flux difference $\Delta \psi$ is normalised by $(B_0 c / \omega_{pi})$ and 
then a unity is subtracted to start from zero.
Time is normalised by the ion cyclotron frequency $\omega_{ci} = e B / m_i$.
Here $\alpha=1.2$.}
\end{figure}

The next question we consider is: how does the time 
evolution of the reconnected flux vary with the ion-electron mass ratio?
Such question historically was relevant because of the inability of
performing realistic ion-electron mass ratio (1836) numerical simulation, due to lack
of computational resources. Although within our reach computationally, we do not 
show here results $m_i / m_e > 100$ because the total energy conservation
error (which is defined as $(E(\omega_{pe} t = 250) - E(\omega_{pe} t = 0))/E(\omega_{pe} t = 0)$
and is indicative
of the code accuracy) starts to deteriorate to values of circa 10\% for $m_i / m_e = 400$, while
for $m_i / m_e = 100$   it is 0.04\% (both for $\alpha=1.2$).
In order to be able to compare our results with the previous work \cite{hesse99},
when varying $m_i / m_e $, we accordingly adjust number of spatial grid points 
and total time step. Such adjustments insure that spatial scale of the simulation box,  is 
$L_x = L_y = 4 c / \omega_{pi}$, and 
the time scale, $\omega_{ci} t = 25$.
Thus when setting $m_i / m_e =1, 9, 25, 64$ and $100$,
Accordingly, the system size is adjusted to $40, 120, 200, 320$ and $400 \Delta$.
The global reconnection size is fixed at $L = 200 \Delta$.
Ion cyclotron frequency for each case is 
defined using by the magnetic intensity at the boundary.

In 2D the magnetic flux function can be defined as
 $\psi = - \int B_x dy = \int B_y dx$.
In our simulation, X-point is located at the centre of system $(x, y) = (0, 0)$, 
while  O-points are at $(x, y) = (0, -2)$ and $(0, 2)$. 
Note that spatial coordinates here are normalised by ion skin depth, and $L_x = L_y = 4 c / \omega_{pi}$.
Therefore we can use the same definition of the 
reconnected flux as in the case tearing mode-unstable Harris current sheet \cite{hesse99}.
Figure \ref{fig:mass_depend} shows time evolution of the magnetic flux difference 
between O and X points (reconnected flux) 
for different ion-to-electron mass ratios, $m_i / m_e = 1, 9, 25, 64$ and $100$.
We gather from this graph that time dynamics of the reconnected flux
does not depend on $m_i / m_e$ when $m_i / m_e \gg 1$ and that reconnection
 is fast. In fact, the time derivative 
of the reconnected flux is the reconnection rate. Thus, the conclusion is that
the reconnection rate is independent of the mass ratio (when $m_i / m_e \gg 1$). As with above
conclusion (in previous subsection) that the reconnection electric field is generated by 
the electron pressure tensor off-diagonal terms; again similarity with the tearing-unstable
Harris current sheet holds, i.e. \citet{hesse99} came to the same conclusion in their case.

As a further test, we performed a numerical run with $m_i / m_e = 1$ (case of
electron-positron plasma). One of the main conclusions of \citet{birn01} was that
as long as Hall term is included, the reconnection is fast. i.e.
when electron and ion dynamics can be distinguished. They showed that slow
reconnection occurs only in the case of single fluid resistive 
MHD (in which there is no
distinction in the electron-ion dynamics). However, in two-fluid MHD or
PIC simulation it is possible to switch off the Hall term by setting $m_i / m_e = 1$
as this will make electron-ion dynamics indistinguishable. The result is given by 
the dotted line in Fig. (\ref{fig:mass_depend}). It can be clearly seen that
the amount of reconnected flux grows very slowly with time, indicating that
the reconnection is slow, as expected.

We propose the following conjecture to explain why the reconnection is fast when
the Hall term is included. Inclusion of the latter means that
in the reconnection inflow magnetic field is frozen into {\it electron} fluid.
As it was previously shown in \citet{th07} (see their Figs.(7) and (11))
speed of electrons, during the reconnection peak time, is
 at least 4-5 times greater than that of ions. This means that electrons can
 bring in / take out the magnetic field attached to them into / away from 
 the diffusion region
 much faster than in the case of single fluid MHD which 
 does not distinguish between
 electron-ion dynamics. In fact, in Fig.(2) the amount of
 reconnected flux attained by $ \omega_{ci} t=25$ 
 in the cases of $m_i / m_e \gg 1$ and $m_i / m_e = 1$
 has the same ratio ($0.11/0.03 \approx 4$)  as is
 the ratio of electron and ion speeds ($\approx 4-5$).
 
 It should be mentioned that although the importance of the 
 Hall term for providing the fast reconnection has been firmly established,
some results indicate \cite{bb07,dk07,hz07} that the fast reconnection
 without it is still possible. This indicates that the issue of
 which physical factor(s) uniquely guarantee the fast rate is still open.

\subsection{Energy budget of the reconnection process}

The next question we address is: what is the exact energy 
budget of the reconnection process, i.e. in which proportion initial magnetic 
energy is converted into other forms of energy?

\begin{figure}
\includegraphics[scale = 0.5]{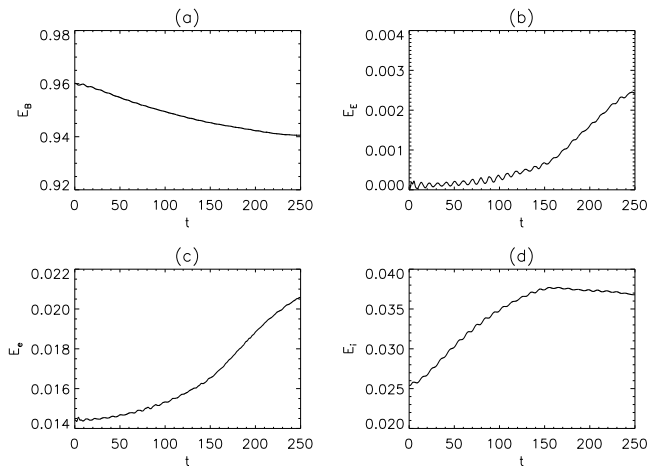}
\caption{\label{fig:conv_a120}
Time evolution of (a) magnetic field energy, 
(b) electric field energy, relativistic (c) electron and 
(d) ion kinetic energies of the whole system for $\alpha = 1.20$.
These energies are normalised by the initial total energy.
Time is normalised by the electron plasma frequency $\omega_{pe}$.}
\end{figure}

\begin{figure}
\includegraphics[scale = 0.5]{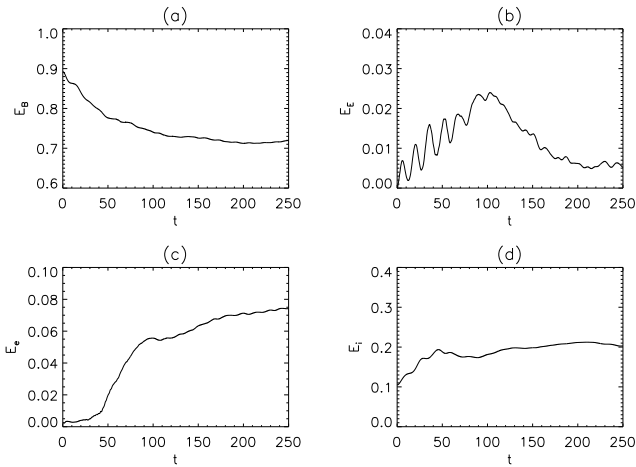}
\caption{\label{fig:conv_a224}
The same as in Fig.~\ref{fig:conv_a120}, but for $\alpha = 2.24$.}
\end{figure}

Figure~\ref{fig:conv_a120} shows time evolution 
of (a) magnetic field energy, (b) electric field energy, 
relativistic (c) electron and (d) ion kinetic energies of
 the whole system for $\alpha = 1.20$.
According to the previous results \cite{th07}, in this case
the normalised reconnection rate peaks at $E_z = 0.11$ at time $ \omega_{pe} t = 170$.
Initially magnetic field energy is dominant, which constitutes $96$\%
of the total energy of system.
The rest $4$\%  goes to 
 the initial electron and ion kinetic energies
 because we impose a non-zero current $j_z$ at $t=0$ 
 according to Eq.(\ref{eq:init_j}) (as $\alpha > 1$).
We gather from  Fig.~\ref{fig:conv_a120}(a) that as the reconnection proceeds 
magnetic field energy is converted into other forms of energy.
As it is also stated in \citet{th07} at $\omega_{pe} t=250$, which corresponds to 
about 1.25 Alfv\'en times, $(0.96-0.94)/0.96 = 2$ \% of the initial magnetic energy is released.
Here we explore partition into which other forms this energy goes into.
Panels (b)-(d) in Fig.~\ref{fig:conv_a120} show that all other forms of energy increase
as time progresses. In particular, electric field energy that starts from
zero, attains value of 0.0024, i.e. $(0-0.0024)/(0.96-0.94)=12$\% of the consumed magnetic energy.
One can conjecture that ultimately this energy will go into particle kinetic energy
(as particles would be easily accelerated by electric fields).
Relativistic kinetic energy of electrons attains $(0.0205-0.0145)/(0.96-0.94)=30$\%,
while the same for ions  $(0.037-0.0255)/(0.96-0.94)=58$\%.
Given that electrons have small inertia and thus 
are more influenced by the electric field, we conjecture that within few Alfv\'en times
electron-ion kinetic energy partition (as the percentage of consumed magnetic energy)
will be roughly 40\% - 60\%. 
It should be mentioned that, in general, it is not easy for an electric 
field to accelerate particles, unless it is 
parallel to the magnetic field or the magnetic field is small, such that the 
particles are non-adiabatic (loosely tied to magnetic field lines). The 
latter condition is more stringent for electrons.
Movies of electric field in our numerical simulation
show complicated, and yet coherent, oscillatory patterns.
We have not performed detailed analysis of identification of nature of these
waves, but based on previous experience, the case without the 
guide field considered here,
would support excitation of whistler waves as the X-point collapses.
In turn, we conjecture that the whistler waves are ultimately responsible for
the particle acceleration. More detailed analysis of this is needed in the future.
Returning to the issue of the established 40\% - 60\% energy partition, \citet{emslie04} showed that the energy
of accelerated electrons is comparable to that of accelerated ions. However,
they admit to large uncertainties in the ion energy spectrum. Despite of this,
our simulation results broadly agree with the solar flare observations \cite{emslie04}.

Previously we also considered strongly stressed X-point ($\alpha = 2.24$) \cite{th07}. In this case 
in 1.25 Alfv\'en times, $(0.9-0.72)/0.9 = 20$\% of the initial magnetic energy is converted
into other forms of energy (this is equivalent of 
$(0.9-0.72) = 18$\% of the initial {\it total} energy; and as we saw in the 
weakly stressed case, the difference between the two is negligible. It is only with 
the increase of $\alpha$ the difference between initial {\it magnetic} energy and 
initial {\it total} energy becomes noticeable, because stronger initial currents 
(i.e. initial kinetic energy of particles) need to be imposed according to Eq.(\ref{eq:init_j})). 
This 18\% decrease in the magnetic energy is also corroborated in panel (a) in Fig.~\ref{fig:conv_a224}.
Exact break down (partition) of the latter is as follows (based on panels (b)-(d)):
 electric field energy that starts from
zero, peaks and then settles at 0.006, i.e. $(0-0.006)/(0.9-0.72)=3$\% of the consumed magnetic energy.
Relativistic kinetic energy of electrons attains $(0.075-0.002)/(0.9-0.72)=41$\%,
while the same for ions  $(0.2-0.1)/(0.9-0.72)=56$\%.
As in the weakly stressed case, within 1.25 Alfv\'en times,
electron-ion kinetic energy partition (as the percentage of total energy, 
which for a solar flare would be the total energy released by flare)
is roughly 40\% - 60\%. This again is in accord to solar flare observations \citet{emslie04}.

\subsection{Properties of velocity distribution of the accelerated particles}

The final question we address is: are there any anisotropies in the velocity distribution of the 
accelerated particles? This question naturally comes to one's mind due to a recent study of
\citet{kontar06}, who surprisingly found 
near-isotropic electron distributions in solar flares, which contrast 
strongly with the expectations from the standard model 
that invokes strong downward beaming, including the collisional thick-target model.

\begin{figure}
\includegraphics[scale = 0.5]{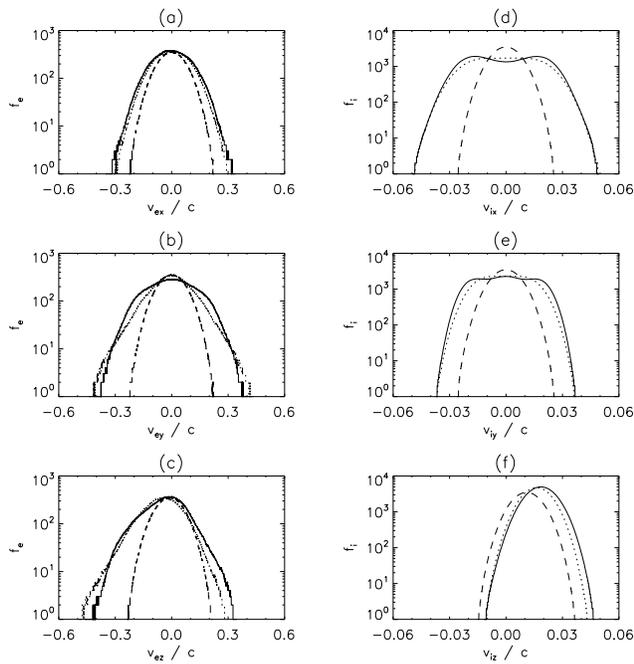}
\caption{\label{fig:vd_a120}
(a-c) Electron and (d-f) ion velocity distribution 
functions in $x$, $y$ and $z$ directions near the 
current sheet at the initial stage $t = 0$ (dashed line), 
the peak reconnection stage $t = 170$ (dotted line) and 
the final simulation time $t = 250$ (solid line) for $\alpha = 1.20$.
As in Ref.\cite{th07}, here data are 
produced using the region of the current sheet
 ($-2 (c/ \omega_{pe}) \leq x \leq 2(c/ \omega_{pe}), -8(c/ \omega_{pe}) \leq y \leq 8(c/ \omega_{pe})$).
$f_e$ and $f_i$ are the number of electrons and ions, respectively.
Velocity and time are normalised by light speed $c$ 
and $\omega_{pe}$, respectively.}
\end{figure}

\begin{figure}
\includegraphics[scale = 0.5]{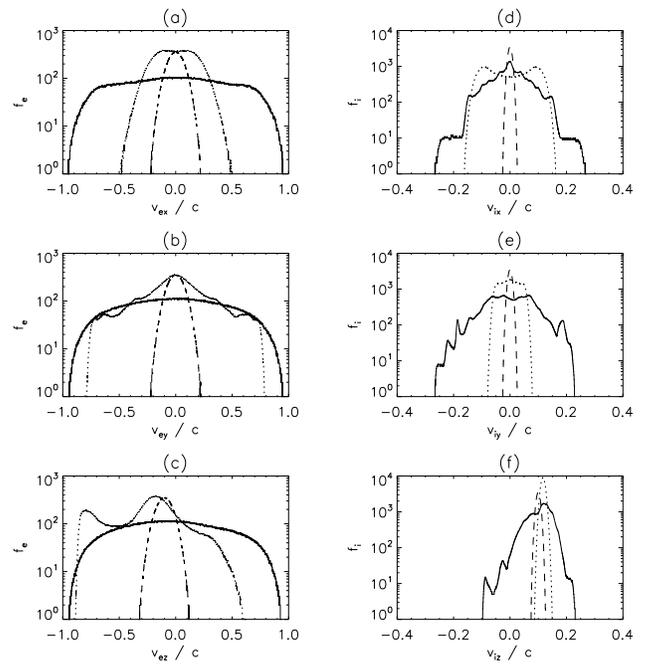}
\caption{\label{fig:vd_a224}
(a-c) Electron and (d-f) ion velocity distribution 
functions in $x$, $y$ and $z$ directions near the 
current sheet at the initial stage $t = 0$ (dashed line), 
the peak reconnection time $t = 45$ (dotted line) and 
the final simulation time $t = 250$ (solid line) for $\alpha = 2.24$.
Data are produced using the region of the current sheet
($-1 (c/ \omega_{pe}) \leq x \leq 1 (c/ \omega_{pe}), -16 (c/ \omega_{pe})\leq y \leq 16 (c/ \omega_{pe})$).
The normalisation is same as in Fig~\ref{fig:vd_a120}.
}
\end{figure}

As in \citet{th07}, here we consider two cases of the weakly and strongly stressed X-point. 
The results are shown in Figures \ref{fig:vd_a120} and \ref{fig:vd_a224}. The following observations can be
 made:

In the weakly stressed case, for electrons we see appearance of 
super-thermal electrons towards the end of simulation time (shortly 
after the peak of time-transient reconnection) mostly in $y$ and $z$ velocity distribution function 
components. 
Dynamics of the flows and currents is presented in detail in Ref.\cite{th07}. 
Here we only mention that the reconnection inflow is in $x$ direction, while the 
outflow is in $y$ direction. Thus based on panels (a)-(c) in Fig.(\ref{fig:vd_a120}) we gather
that accelerated electrons (focus on solid and dotted curves) are due to reconnection outflow (in reconnection plane) 
as well as out-of-plane flow (which is triggered by the out-of-plane electric
field generated at the magnetic null). For ions, at later 
stages of the reconnection, in panels (d)-(f) in 
Fig.(\ref{fig:vd_a120}) we see (focus on solid and dotted curves)  a 
superposition of two Maxwellian distributions in both reconnection 
inflow (along $x$) and outflow (along $y$). These seem to be created 
by reconnection flow dynamics. In $z$-direction we see a shifted 
(also somewhat broadened by the heating) Maxwellian, which is due 
to out-of-plane ion beam (localized current).

We gather from panels (a)-(b) in Fig.(\ref{fig:vd_a224}) (focus now only on solid curves) that in the 
strongly stressed X-point case, in about one Alfv\'en time, super-thermal 
electrons show a full isotropy in all three spatial directions of 
the velocity distribution. In solar flare observations \citet{kontar06} report that 
electron distributions are also nearly isotropic, which seems to contradict to 
what is expected from the standard model flare models that 
invoke strong downward beaming of electrons. In this respect, 
the match of our simulation results with the flare observations 
seem encouraging, in that stressed X-point collapse seems to be a 
viable mechanism acting during solar flares. For ions (panels (d)-(f) 
in Fig.(\ref{fig:vd_a224})) behaviour is not so much different from 
the weakly stressed case (panels (d)-(f) in Fig.(\ref{fig:vd_a120})), 
except for much higher velocities attained and distribution functions 
modified by kinetic, wave-particle interaction instabilities. 
The latter can be judged by sign changes in the slope of the distribution 
function, which can only occur when waves and particles exchange 
energy and momentum.

\section{Conclusions}

By and large, the present work closes our initial study of 
stressed X-point collapse in the collisionless regime started in 
Ref.\cite{th07}, by bridging gaps in the understanding of key physical 
aspects. The main findings can be listed as following:

(i) despite significant differences of the initial setup between 
tearing unstable Harris current sheet \cite{pritchett01} and stressed 
X-point considered here, in both cases source of the reconnection 
out-of-plane electric field at the magnetic null is provided by 
off-diagonal terms of the electron pressure tensor.

(ii) we find that when $m_i / m_e \gg 1$ reconnection rate is independent of the 
ion-electron mass ratio and it is fast, which is also witnessed by \citet{hesse99}. 
However, when electron-ion mass ratio is unity, i.e. the Hall term 
is switched off, we show that reconnection rate is indeed slow. 
This broadly agrees with the results of \citet{birn01} (However see also Refs.\cite{bb07,dk07,hz07}
for alternative view).
When the Hall physics is included, we also conjecture that the reconnection is fast  
because the magntic field (being frozen into electron fluid, which
moves significantly faster than ion fluid, as shown in Ref.\cite{th07}) is 
transported in and out of the
diffusion region much faster than in the case of single fluid resistive MHD.
We show that the amount of
 reconnected flux attained by $ \omega_{ci} t=25$
 in the cases of $m_i / m_e \gg 1$ and $m_i / m_e = 1$
 has the same ratio ($\approx 4$)  as is
 the ratio of electron and ion speeds ($\approx 4-5$).

(iii) we find that within one Alfv\'en time, roughly $\sim 40$\% of 
the 
initial total energy (which is mostly stored in the magnetic field) 
is converted into the kinetic energy electrons, and somewhat 
more than half ($\sim 60$\%) into kinetic energy of ions. 
In solar flare observations a similar behaviour is seen \cite{emslie04}.

(iv) When X-point is stressed {\it strongly}, in about one Alfv\'en time, a full 
isotropy in all three spatial directions of the velocity distribution is 
seen for super-thermal electrons. Again similar behaviour is reported 
in  solar flare observations \cite{kontar06}.

Resuming aforesaid, it seems that collisionless, stressed X-point 
collapse is a viable mechanism for solar flares. Also, its behaviour 
is remarkably similar to tearing unstable Harris current sheet which 
is thought to be more relevant for the Earth geomagnetic tail and 
generally to magnetospheric applications.

We close this paper with some words of caution: 

(i) realising that 
size of the numerically simulated area $40 \times 40$
electron skin depths ($c / \omega_{pe}=0.16804$ m in solar corona), i.e. 
$6.7\times 6.7$ m is a tiny proportion of the actual solar flare site
which can be tens of Mm. Thus, {\it a priori} it is 
not clear whether this mechanism is suitable 
to accelerate enough electrons in a much larger volume.

(ii) whether e.g. the obtained electron-ion kinetic energy partition 
(as the percentage of total released energy)
being roughly 40\% - 60\%, would hold for a different mass ratio?
We remind the reader that except in subsection III.B 
the ion-electron mass ratio was kept
constant at $m_i/m_e=100$. We can only conjecture that 
this  40\% - 60\% partition 
holds for a different mass ratio, 
because we also found that the amount of reconnected flux versus time
graph (Fig.~2) showed no mass ratio dependence.

\begin{acknowledgments}
This research was supported by the United Kingdom's Science and 
Technology Facilities Council (STFC).
\end{acknowledgments}


\end{document}